# Analyzing Features for the Detection of Happy Endings in German Novels

Fotis Jannidis, Isabella Reger, Albin Zehe, Martin Becker, Lena Hettinger, Andreas Hotho

**Abstract**

With regard to a computational representation of literary plot, this paper looks at the use of sentiment analysis for happy ending detection in German novels. Its focus lies on the investigation of previously proposed sentiment features in order to gain insight about the relevance of specific features on the one hand and the implications of their performance on the other hand. Therefore, we study various partitionings of novels, considering the highly variable concept of "ending". We also show that our approach, even though still rather simple, can potentially lead to substantial findings relevant to literary studies.

**Introduction**

Plot is fundamental for the structure of literary works. Methods for the computational representation of plot or special plot elements would therefore be a great achievement for digital literary studies. This paper looks at one such element: happy endings.

We employ sentiment analysis for the detection of happy endings, but focus on a qualitative analysis of specific features and their performance in order to gain deeper insight into the automatic classification. In addition, we show how the applied method can be used for subsequent research questions, yielding interesting results with regard to publishing periods of the novels.

**Related Work**

One of the first works was on folkloristic tales, done by Mark Finlayson, who created an algorithm capable of detecting events and higher-level abstractions, such as villainy or reward (Finlayson 2012). Reiter et al., again on tales, identify events, their participants and order and use machine learning methods to find structural similarities across texts (Reiter 2013, Reiter et al. 2014).

Recently, a significant amount of attention has been paid to sentiment analysis, when Matthew Jockers proposed emotional arousal as a new "method for detecting plot" (Jockers 2014). He described his idea to split novels into segments and use those to form plot trajectories (Jockers 2015). Despite general acceptance of the idea to employ sentiment analysis, his use of the Fourier Transformation to smooth the resulting plot curves was criticized (Swafford 2015, Schmidt 2015).

Among other features, Micha Elsner (Elsner 2015) builds plot representations of romantic novels, again by using sentiment trajectories. He also links such trajectories with specific characters and looks at character co-occurrences. To evaluate his approach, he distinguishes real novels from artificially reordered surrogates with considerable success, showing that his methods indeed capture certain aspects of plot structure.

In previous work, we used sentiment features to detect happy endings as a major plot element in German novels, reaching an F1-score of 73% (Zehe et al. 2016).

**Corpus and Resources**

Our dataset consists of 212 novels in German language mostly from the 19th century[1]. Each novel has been manually annotated as either having a happy ending (50%) or not (50%). The relevant information has been obtained from summaries of the Kindler Literary Lexikon Online[2] and Wikipedia. If no summary was available, the corresponding parts of the novel have been read by the annotators.

Sentiment analysis requires a resource which lists sentiment values that human readers typically associate with certain words or phrases in a text. This paper relies on the NRC Sentiment Lexicon (Mohammad and Turney 2013), which is available in an automatically translated German version[3]. A notable feature of this lexicon is that besides specifying binary values (0 or 1) for negative and positive connotations (2 features) it also categorizes words into 8 basic emotions (anger, fear, disgust, surprise, joy, anticipation, trust and sadness), see Table 1 for an example. We add another value (the polarity) by subtracting the negative from the positive value (e.g. a word with a positive value of 0 and a negative value of 1 has a polarity value of -1). The polarity serves as an overall sentiment score, which results in 11 features.

*Table 1: Example entries from the NRC Sentiment Lexicon*

| Word/Dimension | verabscheuen (to detest) | bewundernswert (admirable) | Zufall (coincidence) |
|---|---|---|---|
| Positive | 0 | 1 | 0 |
| Negative | 1 | 0 | 0 |
| Polarity | -1 | 1 | 0 |
| Anger | 1 | 0 | 0 |
| Anticipation | 0 | 0 | 0 |
| Disgust | 1 | 0 | 0 |
| Fear | 1 | 0 | 0 |
| Joy | 0 | 1 | 0 |
| Sadness | 0 | 0 | 0 |
| Surprise | 0 | 0 | 1 |
| Trust | 0 | 1 | 0 |

**Experiments**

The goal of this paper is to investigate features that have been used for the detection of happy endings in novels in order to gain insight about the relevance of specific feature sets on the one hand and the implications of their performance on the other hand. To that end, we adopt the features and methods presented in Zehe et al. (2016). The parameters of the linear SVM and the partitioning into 75 segments are also adopted from this paper.

---

[1] Source: https://textgrid.de/digitale-bibliothek
[2] www.kll-online.de
[3] http://saifmohammad.com/WebPages/NRC-Emotion-Lexicon.htm

*Features.* Since reliable chapter annotations were not available, each novel has been split into 75 equally sized blocks, called *segments*. For each lemmatized word, we look up the 11 sentiment values (including polarity, see above). Then, for each segment, we calculate the respective averages, resulting in 11 scores per segment. We group those 11 scores into one feature set.

*Qualitative Feature Analysis.* As our corpus consists of an equal number of novels with and without happy ending, the random baseline as well the majority vote baseline amount to 50% classification accuracy.

Since we assumed that the relevant information for identifying happy endings can be found at the end of a novel, we first used the sentiment scores of the final segment ($f_{d,n}$) as the only feature set, reaching an F1-score of 67%.

Following the intuition that not only the last segment by itself, but also its relation to the rest of the novel are meaningful for the classification, we introduced the notion of *sections*: the last segment of a novel constitutes the *final section*, whereas the remaining segments belong to the *main section*. Averages were also calculated for the sections by taking the mean of each feature over all segments in the section. To further emphasize the relation between these sections, we added the differences between the sentiment scores of the final section and the average sentiment scores over all segments in the main section. However, this change did not influence the results.

This led us to believe that our notion of an "ending" was not accurate enough, as the number of segments for each novel and therefore the boundaries of the final segment have been chosen rather arbitrarily. To approach this issue, we varied the partitioning into main and final section so that the final section can contain more than just the last segment.

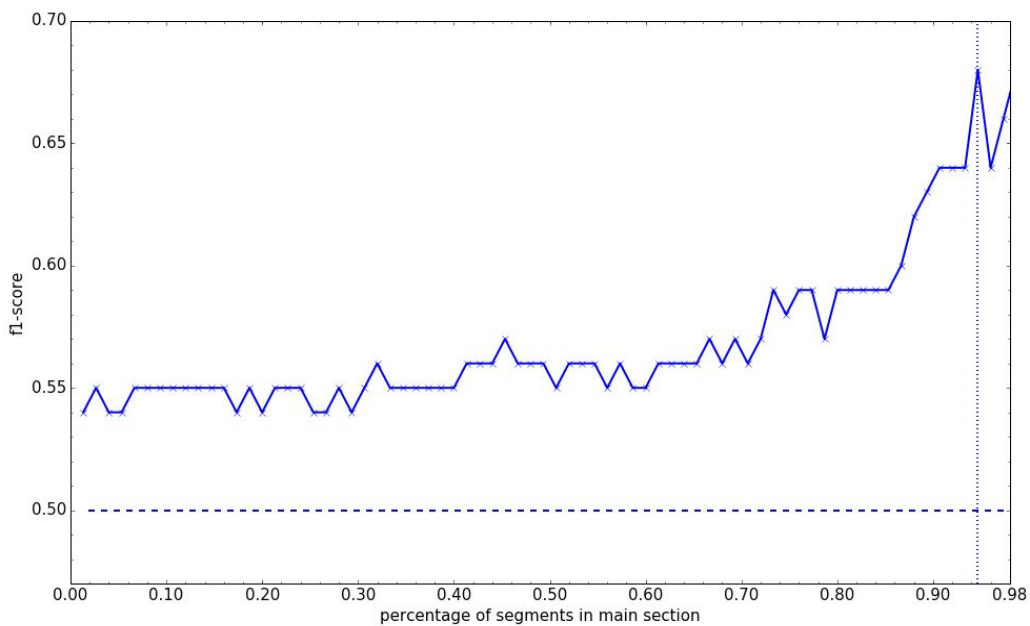

**Figure 1**: Classification F1-score for different partitionings into main and final section. The dashed line represents a random baseline, the dotted line shows where the maximum F1-score is reached.

Figure 1 shows that classification accuracy improves when at least 75% of the segments are in the main section and reaches a peak at about 95% (this means 4 segments in the final section and 71 segments in the main section, for a total of 75 segments). With this partitioning strategy, we improve the F1-score to 68% using only the feature set for the final section ($f_{d,final}$) and reach an F1-score of 69% when also including the differences to the average sentiment scores of the main section ($f_{d,\,main-final}$).

Since adding the relation between the main section and the final section improved our results in the previous setting, we tried to model the development of the sentiments towards the end of the novel in a more profound way. For example, a catastrophic event might happen shortly before the end of a novel and finally be resolved in a happy ending. To capture this intuition, we introduced one more section, namely the *late-main section,* which focuses on the segments right *before* the final section, and used the difference between the feature sets for the late-main and the final section as an additional feature set ($f_{d,\,late-final}$). Using those three feature sets, the classification of happy endings reaches an F1-score of 70% and increases to 73% when including the feature set for the final segment.

*Table 2*: Classification F1-score for the different feature sets

| **Features** | **Results** |
| --- | --- |
| 1) Final segment feature set | 67% |
| 2) Final segment feature set and difference to main section | 67% |
| 3) Final section feature set with final section of length 4 | 68% |
| 4) Feature set 3 and difference to main section | 69% |
| 5) Feature set 4 and difference between late-main section and final section | 70% |
| 6) Feature set 5 and final segment feature set | 73% |

Table 2 summarizes these results and shows that the addition of each feature set leads to small improvements, amounting up to a F1-score of 73%. While we saw that the classification performs best when the final section consists of 4 segments, we also observed that quite a few novels could be correctly classified with several different partitionings. On the other hand, some novels could not be predicted correctly with any choice of partitioning. An example is *Twenty Thousand Leagues Under the Sea* by Jules Verne which evidently has a happy ending with clearly identifiable boundaries, but an extremely short one, consisting only of about 250 words. These observations show that the notion of a novel's "ending" is highly variable and can differ considerably from text to text.

*Correlation with Publication Dates.* This raises the question whether we can use the sensitivity of our approach to this kind of variability in order to better understand the characteristics of the novels in our corpus. As an example, we studied whether different section partitionings are in any way correlated with the publication date of a novel. In order to keep the results as interpretable as possible, we focused on one single feature set: the sentiment scores of the final section. In a first attempt, we divided our corpus into four subgroups, distinguishing novels published before 1830 (65 novels), between 1831 and 1848 (31 novels), between 1849 and 1870 (29 novels) and after 1871 (87 novels). This split resulted in similarly sized portions and did not yield a strong bias towards happy/unhappy endings in any period.

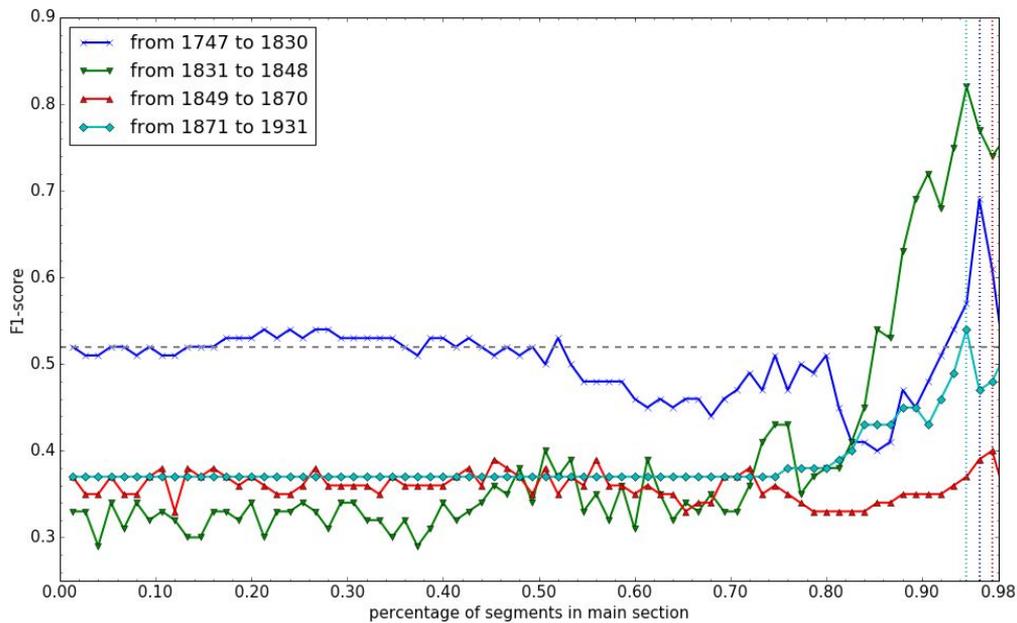

**Figure 2**: Classification F1-score for different partitionings into main and final section. Each line denotes novels from a different time period. The dashed line represents the random baseline for the time period starting from 1871. Random baselines for the other periods yield slightly worse results and are omitted. The dotted lines show where the maximum F1-score is reached for the respective time periods.

Figure 2 shows that the best classification is again obtained when about 95 - 98% of the segments are in the main section, regardless of the time period. Therefore, the best section split point is not correlated with the publication date of a novel. What is striking, however, is the fact that the novels published after 1848 yield considerably lower scores than the novels published before that year, mostly even below the baseline. This indicates a correlation between publication date and automatic classification quality, i.e. novels published before the period of Realism are more easily classifiable in terms of having a happy ending than realistic novels. A possible explanation is that many novels of that earlier period are more schematically structured.

We are aware that the number of novels for each of the time spans is rather small, so that those findings can only be regarded as exploratory insights. Nevertheless, these preliminary results show that the automatic detection of happy endings, even with only one rather simple feature set, can uncover dependencies to other properties of novels that are highly interesting for literary studies.

**Conclusion and Future Work**

The automatic detection of happy endings as a major plot element of novels is a valuable step towards a comprehensive computational representation of literary plot. Our experiments show that different features based on sentiment analysis can predict happy endings in novels with varying but reasonable quality. Even though our approach is still rather simple, we showed that it can potentially lead to substantial insights for literary scholars.

Future work may cover improving our classification by accounting for the high variability of endings in novels and may also include further leveraging our approach to study the characteristics of different novel collections in-depth.

**Zehe, Albin / Becker, Martin / Hettinger, Lena / Hotho, Andreas / Reger, Isabella / Jannidis, Fotis** (2016): "Prediction of Happy Endings in German Novels", in: *Proceedings of the Workshop on Interactions between Data Mining and Natural Language Processing 2016*.